**Relaxation Dynamics of Superconducting/Normal Interfaces in the Intermediate State of a type-I superconductor**

Mengju Yuan[1], Yugang Zhang[1], Ying Zhu[1], Jingchun Gao[1], Aifeng Wang[1], Mingquan He[1], Jun-Yi Ge[2,†], Yisheng Chai[1,§]

[1]*Low Temperature Physics Laboratory, College of Physics & Center of Quantum Materials and Devices, Chongqing University, Chongqing 401331, China*

[2]*Materials Genome Institute, Shanghai University, Shanghai 200444, China*

Contact author:

[†] Contact author: junyi_ge@t.shu.edu.cn

[§] Contact author: yschai@cqu.edu.cn

The dynamics of phase boundaries, such as superconducting/normal (S/N) interfaces in type-I superconductors, are typically obscured in conventional magnetic measurements, which are strongly influenced by surface barriers and global flux processes. Here, we employ the complex ac magnetostrictive coefficient as a sensitive probe of the bulk dynamics of S/N interfaces in the intermediate state of lead. The ac magnetostrictive coefficient exhibits an unusual complex response, with a negative real component and a positive imaginary component. Frequency-dependent measurements show that the magnitude of the real component decreases monotonically with increasing frequency, whereas the imaginary component first increases, reaches a broad maximum, and then decreases. In contrast, ac susceptibility displays a conventional Debye-type magnetic relaxation with a different phase relation between its real and imaginary components. We propose that the magnetostrictive response arises from ac-field-driven relaxation and rearrangement of metastable S/N domain configurations toward states of lower free energy. These results reveal a distinct interfacial relaxation channel in the intermediate state and establish the complex ac magnetostrictive coefficient as a sensitive probe of nonequilibrium S/N interface dynamics.

# I. INTRODUCTION

The dynamics of interfaces and domain walls govern the functional properties of a vast array of quantum materials, representing an important theme in condensed matter physics. From the sliding motion of charge density waves [1–3] and the switching of ferroelectric domains [4–7] to the dynamics of topological spin textures such as skyrmions [8–11], the collective motion of mesoscale structures often dictates their macroscopic response. Yet, deciphering their intrinsic relaxation pathways, energy landscapes, and dissipation mechanisms remains a formidable challenge. Type-I superconductors offer a particularly clean platform to explore such physics through the intermediate state (IS), where superconducting (S) and normal (N) domains form a self-organized, spatially modulated state separated by mobile S/N interfaces [12–14]. As shown in the phase diagram [Fig. 1(a)], this state emerges in finite samples with a demagnetization factor $n$ when the applied magnetic field ($H$) exceeds $H_c(1\text{-}n)$ (where $H_c$ is the thermodynamic critical field), and occupies the field range between the completely diamagnetic Meissner state and the normal state.

Despite decades of study of the static morphology of these domains, revealing intricate patterns such as laminar stripes, flux tubes, and suprafroth structures [14–17], their dynamical response under time-varying fields [18] remains difficult to isolate from boundary-related effects. Conventional probes, primarily ac magnetic susceptibility ($\chi = \chi' - i\chi''$), are strongly influenced by surface barriers, geometric constraints, and dissipative flux penetration in finite samples [19–24]. Consequently, the intrinsic motion and rearrangement of S/N interfaces can be masked by the global magnetic response. In particular, whether metastable S/N domain configurations can relax and reorganize under an ac drive, and what characteristic timescales govern such processes, remain poorly understood. This limitation has prevented a complete understanding of the nonequilibrium energy landscape governing these prototypical modulated phases.

A complementary approach to these dynamics is provided by the complex ac magnetostrictive coefficient, $(\mathrm{d}\lambda/\mathrm{d}H)_{ac} = \mathrm{d}\lambda'/\mathrm{d}H + i\mathrm{d}\lambda''/\mathrm{d}H$, ($\lambda = \Delta L/L$, $L$ is the dimension of the sample), which can be detected with high sensitivity using a composite magnetoelectric technique [25–27]. Unlike conventional magnetostriction, which measures the field-induced change in sample dimensions, complex ac magnetostriction probes the differential strain response to a weak oscillating field and resolves both its in-phase and dissipative components. This technique has proven sensitive to slow collective dynamics, including magnetic-domain and domain-wall motion in ferromagnets [27] and vortex entry, pinning, and dissipation in type-II superconductors [26]. These studies establish complex ac magnetostriction as a complementary probe of collective degrees of freedom coupled to lattice strain. It is therefore well suited for investigating the intermediate state of a type-I superconductor, where the relevant mesoscale dynamics arise from the motion and rearrangement of extended S/N interfaces rather than quantized vortices.

Here we extend this approach to the intermediate state of elemental Pb and examine the dynamics of its S/N domain structure by combining complex ac magnetostriction with ac susceptibility, field- and temperature-dependent measurements. We first establish that the finite magnetostrictive response is confined to the intermediate-state region bounded by the Meissner and normal states and that pronounced hysteresis accompanies the evolution of the S/N domain structure, demonstrating the importance of metastability and barrier-controlled domain rearrangement. We then show that the frequency-dependent magnetostrictive response possesses a characteristic relaxation timescale but an unusual phase relation, with $d\lambda'/dH < 0$ and $d\lambda''/dH > 0$, which cannot be reduced to a conventional single-component Debye response. Its behavior is also distinct from the Debye-like relaxation observed in ac susceptibility, indicating that the two measurements emphasize different dynamical channels. We interpret the magnetostrictive response in terms of ac-field-assisted relaxation and redistribution of metastable S/N configurations within a multidimensional free-energy landscape. The results therefore provide a strain-based view of nonequilibrium dynamics in the intermediate state and show that complex ac magnetostrictive coefficients can complement conventional magnetic probes by selectively accessing interfacial rearrangements coupled to the bulk lattice.

## II. EXPERIMENTAL METHODS

High-purity lead (Pb, 99.9998%) ingot purchased from Alfa Aesar was used in this study. The Pb crystal has dimensions of 1.7 × 1.7 × 0.85 $mm^3$ that we define a Cartesian coordinate system with the thickness direction along *c*-axis and in-plane directions as *a*- and *b*-axes [Fig. 1(b)]. The demagnetization factors *n* is estimated to be 0.25 (in-plane) and 0.50 (out-of-plane) respectively [28]. The temperature-dependent resistivity was measured via a standard four-probe method. The ac susceptibility and dc magnetization were measured using a 7 T Magnetic Property Measurement System. To probe the ac magnetostrictive response, a composite magnetoelectric (ME) structure was fabricated by bonding the Pb samples onto a [001]-cut $0.7Pb(Mg_{1/3}Nb_{2/3})O_3$-$0.3PbTiO_3$ (PMN-PT) single crystal, as shown in the Fig. 1(b). In this method, field-induced strain is converted into an electrical signal $V_{ME}$ with high sensitivity. This $V_{ME}$ is proportional to $(d\lambda/dH)_{ac} = d\lambda'/dH + i d\lambda''/dH$ and has been treated as $(d\lambda/dH)_{ac}$ throughout this paper.

## III. RESULT AND DISCUSSIONS

Figure 2(a) shows the temperature-dependent resistivity of Pb under zero field. Pb exhibits a sharp superconducting transition at $T_c \approx 7.2$ K, consistent with reported values for bulk Pb [29]. Figure 2(b) presents the zero-field-cooled (ZFC) magnetic shielding measured at 10 Oe for $H//c$ and $H//a$. In both orientations, a sharp diamagnetic transition occurs at $T_c$, with a shielding response corresponding approximately to full superconducting screening within experimental uncertainty. These measurements establish the bulk superconducting properties of the sample before examining its intermediate-state dynamics.

Figures 3(a)–3(d) display the magnetic-field and temperature dependence of the $(d\lambda/dH)_{ac}$ for the in-plane orientation with frequency $f = 211$ Hz. In both the Meissner state and the normal state, the magnetostrictive coefficient response remains essentially zero. In contrast, $d\lambda'/dH$ in the vortex lattice phase of type-II superconductors show a quasi-linear dependence on magnetic field [26]. During field sweeps [Figs. 3(a) and 3(b)], the response sharply deviates from zero as the applied field exceeds a lower

threshold $H^*$, signaling the penetration of magnetic flux and the onset of the intermediate state. Within this regime, $d\lambda'/dH$ develops a pronounced negative peak, while $d\lambda''/dH$ simultaneously exhibits a positive peak of opposite sign. As the field reaches the thermodynamic critical field $H_c$, the sample enters the normal state, and the response vanishes. Consistent dynamic features are observed during temperature sweeps [Figs. 3(c) and 3(d)], where the intermediate state is robustly bounded between $T_c$ and a lower characteristic temperature corresponding to the intermediate state to the Meissner state transition. This unique sign-reversed response differs markedly from the conventional magnetostrictive dissipation observed in the vortex liquid regime of type-II superconductors using the same technique [26], reflecting the coexistence and dynamics of S/N domains in type-I superconductor. Furthermore, the field sweeps exhibit pronounced hysteresis between increasing- and decreasing-field branches throughout the intermediate-state region. Such history dependence is consistent with geometric barriers, pinning, and metastable S/N domain configurations, which can follow different evolution pathways during flux penetration and flux expulsion [29].

By extracting these characteristic transition points ($H^*$, $H_c$, $T_c$ and characteristic temperature) from the dynamic responses, we construct the field-temperature ($H$-$T$) phase diagram of Pb in Fig. 3(e). The upper phase boundary, corresponding to the thermodynamic critical field $H_c$, is well described by $H_c(T) = H_c(0)[1-(T/T_c)^2]$, yielding $H_c(0)$ = 789 Oe and $T_c$ = 7.45 K, which are consistent with reported value for bulk Pb [30]. The lower threshold $H^*$ closely follows the theoretical relation $H_c(T)(1-n)$, as expected for the onset of the intermediate state in a finite sample with demagnetization factor $n$. These two boundaries separate the Meissner, intermediate, and normal regimes.

We note that the measurements along the out-of-plane orientation ($H//c$) show qualitatively similar phase boundaries [Fig. 4]. Because the demagnetization factor is larger for this orientation, the intermediate-state region is correspondingly broader. The incomplete disappearance of the magnetostrictive response at low temperatures may additionally reflect geometric barriers and residual pinning, which hinder complete flux expulsion [17].

A striking feature of the intermediate-state response is the opposite sign of the two components, $d\lambda'/dH < 0$ and $d\lambda''/dH > 0$. In most cases, the imaginary component of ac magnetostrictive coefficient is either nearly zero or exhibits finite contribution in the same sign with that of the real part, corresponding to predominantly non-dissipative or dissipative dynamic responses, respectively, as demonstrated in type-II superconductor and ferromagnetic material [26,27]. Interestingly, anomalous phase relations in complex ac responses have previously been observed near first-order magnetic transitions and have been discussed in terms of the coexistence and relaxation of metastable and stable phases [25,27,31]. This analogy is particularly relevant here because the intermediate state of a type-I superconductor consists of coexisting S and N phases, while the pronounced field hysteresis in Figs. 3 and 4 demonstrates strong history dependence of the domain configurations. These observations motivate us to examine whether the unusual magnetostrictive response originates from the relaxation of metastable S/N configurations.

To elucidate the dynamics of the intermediate state, we compare $\chi$ and $(d\lambda/dH)_{ac}$ measured under the same dc and ac magnetic fields over a broad frequency range. Figures 5(a) and 5(b) show the temperature dependence of $\chi$ for $H//a$ at $H = 206$ Oe. With increasing frequency, the magnitude of the in-phase response decreases monotonically, whereas the out-of-phase component first increases, reaches a broad maximum, and subsequently decreases. The corresponding frequency dependence extracted near the intermediate-state transition [Fig. 5(c)] is characteristic of a Debye-like magnetic relaxation [32,33]. In contrast, $(d\lambda/dH)_{ac}$ displays a distinct complex response under otherwise identical conditions. As shown in Figs. 5(d) and 5(e), $d\lambda'/dH$ remains negative whereas $d\lambda''/dH$ remains positive throughout the measured frequency range. The frequency dependence extracted at 6.54 K [Fig. 5(f)] shows that $|d\lambda'/dH|$ decreases monotonically with increasing frequency, while $d\lambda''/dH$ increases to a broad maximum at intermediate frequencies and then decreases. Similar behavior is reproduced for $H//c$, as shown in Fig. 6. Both responses therefore contain characteristic frequencies of order 100 Hz, corresponding to relaxation times on the millisecond scale, but their complex phase relations are distinctly different. This contrast indicates that ac susceptibility and ac magnetostriction probe different aspects of the intermediate-state dynamics.

To rationalize the contrasting responses, we first consider the conventional relaxation observed in ac susceptibility. We introduce a generalized coordinate describing the magnetic response associated with flux redistribution and the accompanying evolution of the S/N domain structure. Its dynamics under a weak ac magnetic field can be described phenomenologically by a driven damped oscillator,

$$m_{\mathrm{w}}\ddot{q} + \beta\dot{q} + \alpha q = F(t) \tag{1}$$

where $m_w$ is the effective interface mass, $\beta$ is the damping coefficient, and $\alpha$ is the restoring force constant. Define $\omega_0^2 = \alpha/m_w$ (resonance angular frequency) and $\omega_{\mathrm{c}} = \alpha/\beta$ (relaxation angular frequency). For ac susceptibility, driving force $F_\chi$ couples to ac magnetic field $H(t) = H_{\mathrm{ac}}e^{i\omega t}$, with driving force amplitude is given by $F_0 = \zeta H_{\mathrm{ac}}$, where ζ is a phenomeno coupling coefficient. In the steady state, $q(t) = q_0 e^{i\omega t}$, with the complex amplitude:

$$q_0(\omega) = \frac{\zeta H_{\mathrm{ac}}}{\alpha - m_{\mathrm{w}}\omega^2 + i\omega\beta} \tag{2}$$

Assuming a linear coupling between the ac magnetization and the interface displacement, $M_{\mathrm{ac}} = \gamma q$ (where $\gamma$ is the magnetic coupling constant), the complex susceptibility is

$$\chi(\omega) = \frac{M_{\mathrm{ac}}}{H_{\mathrm{ac}}} = \frac{\chi_0}{1 - \omega^2/\omega_0^2 + i\omega/\omega_{\mathrm{c}}} \tag{3}$$

where $\chi_0 = \gamma\zeta/\alpha$ is the static susceptibility. Its real and imaginary parts are

$$\chi'(\omega) = \chi_0 \frac{1 - \omega^2/\omega_0^2}{(1 - \omega^2/\omega_0^2)^2 + (\omega/\omega_{\mathrm{c}})^2} \tag{4}$$

$$\chi''(\omega) = \chi_0 \frac{\omega/\omega_{\mathrm{c}}}{(1 - \omega^2/\omega_0^2)^2 + (\omega/\omega_{\mathrm{c}})^2} \tag{5}$$

For Eq. (5), the qualitative spectral response depends on the relative magnitude of the resonance frequency $\omega_0$ and the relaxation frequency $\omega_{\mathrm{c}}$. When $\omega_0 \ll \omega_{\mathrm{c}}$, the system shows a resonance type response: the $\chi''$ exhibits a sharp peak near $\omega \approx \omega_0$, while $\chi'$ crosses from positive to negative with increasing frequency. Such kind of behavior is inconsistent with the experimental results of $\chi$ in Figs. 5(c) and 6(c). In contrast, when $\omega_0 \gg \omega_{\mathrm{c}}$, the response is relaxation like: Eq. (3) reduces to the Debye-type form

$$\chi(\omega) \approx \frac{\chi_0}{1 + i\omega/\omega_c} \quad (6)$$

$\chi'$ decreases monotonically with increasing frequency, whereas $\chi''$ exhibits a broad maximum near $\omega \approx \omega_c$. Figure 7(a) shows the corresponding theoretical curves calculated directly from Eq. (6). This characteristic frequency dependence is fully consistent with the experimental behavior of $\chi$ in Figs. 5(c) and 6(c). The strong damping is plausibly associated with surface and geometric barriers together with dissipative flux-entry/exit processes in the finite type-I sample [17].

The above oscillator description successfully captures the relaxation-like ac susceptibility in the overdamped limit, but it cannot account for the ac magnetostrictive response. In particular, the magnetostrictive data exhibit $d\lambda'/dH < 0$ and $d\lambda''/dH > 0$ throughout the measured frequency range [Figs. 5(f) and 6(f)]. This unusual phase relation suggests that the magnetostrictive response involves a dynamical process distinct from the global magnetic relaxation detected by ac susceptibility. A possible physical picture can be constructed by considering the metastability inherent to the intermediate state. A conceptually similar mechanism has been proposed by A. N. Ulyanov *et al.* for anomalous complex ac susceptibility near first-order magnetic transitions, where ac-field-assisted relaxation between metastable and lower-free-energy states gives rise to an unconventional imaginary response [31].

The S–N transition in a type-I superconductor is thermodynamically first order, while the finite demagnetization factor stabilizes coexistence of the S and N phases over a finite range of applied field. In a real finite sample, the actual S/N domain configuration is governed not only by the bulk free-energy balance but also by the positive S/N interfacial energy, long-range magnetostatic interactions, geometric barriers, pinning, and magnetic-field history. The pronounced hysteresis between increasing- and decreasing-field sweeps observed in Figs. 3 and 4 is therefore consistent with different S/N domain configurations becoming trapped in local minima of a complex free-energy landscape. As schematically illustrated in Fig. 7(b), we denote such a metastable S/N configuration as F1 and a neighboring configuration with lower free energy as F2, separated by a local activation barrier. Under an ac magnetic field, the periodic drive can promote local S/N interfaces to cross these barriers and rearrange toward lower free energy configurations [31] [Fig. 7(c)]. Such rearrangement reduces the excess free energy stored in metastable configurations and changes the strain associated with the S/N domain structure, manifested by opposite sign in imaginary component of ac magnetostrictive coefficient.

This metastable-domain picture also provides a natural interpretation of the characteristic frequency dependence of the ac magnetostrictive response. At low frequencies, the S/N interfaces have sufficient time to respond to the slowly varying magnetic field and to rearrange among accessible configurations, resulting in a large magnitude of the negative $d\lambda'/dH$ and a comparatively small positive $d\lambda''/dH$. As the driving frequency approaches the characteristic rate for barrier-mediated S/N-domain rearrangement, the response increasingly lags behind the ac field and the out-of-phase component $d\lambda''/dH$ grows to a broad maximum. At still higher frequencies, the S/N interfaces can no longer substantially reorganize within one ac cycle, leading to a progressive suppression of both components. The broad frequency dependence is naturally expected because the intermediate state contains a distribution of domain sizes, interface geometries, pinning strengths, and local energy barriers, and therefore a corresponding distribution of relaxation times. Thus, although both $\chi$ and $(d\lambda/dH)_{ac}$ exhibit characteristic relaxation behavior on comparable frequency scales, they emphasize different dynamical channels of the intermediate state: ac susceptibility primarily reflects global magnetic-flux redistribution strongly influenced by surface and geometric barriers, whereas ac magnetostriction is particularly sensitive to strain-coupled rearrangement of metastable S/N domain configurations.

# IV. Conclusion

In summary, we have investigated the dynamic response of the intermediate state in type-I Pb using ac susceptibility and the complex ac magnetostrictive coefficient. The magnetostrictive response appears exclusively in the S/N coexistence regime and shows pronounced hysteresis, together with an unusual complex response characterized by negative $d\lambda'/dH$ and positive $d\lambda''/dH$. Its frequency dependence reveals a characteristic relaxation process that is distinct from the conventional Debye-like relaxation observed in ac susceptibility. These results indicate that ac magnetostriction is particularly sensitive to the relaxation and rearrangement of metastable S/N domain configurations in the intermediate state. Our work provides a complementary approach for probing nonequilibrium interface dynamics in type-I superconductors and may be extended to other systems containing mobile phase boundaries and metastable domain structures.

## ACKNOWLEDGMENTS

Mengju Yuan and Yugang Zhang are contributed equally in this work. We thank Yan Liu at the Analytical and Testing Center of Chongqing University for technical support. This work was supported by the National Natural Science Foundation of China (Grant Nos. 11674347, 11974065, 51725104, 11774399, 11474330, U21A201910), Fundamental Research Funds for the Central Universities (Project No. 2024IAIS-ZX002), the National Key Research and Development Program of China (Grants No. 2018YFA0704200 and No. 2023YFA1406100). Y. S. Chai would like to thank the support from Beijing National Laboratory for Condensed Matter Physics.

## DATA AVAILABILITY

The data that supports the findings of this study are available from the corresponding authors upon reasonable request.

[1] P. Monceau, Adv. Phys. **61**, 325 (2012).

[2] J. Hwang, W. Ruan, Y. Chen, S. Tang, M. F. Crommie, Z.-X. Shen, and S.-K. Mo, Rep. Prog. Phys. **87**, 044502 (2024).

[3] G. Grüner, Rev. Mod. Phys. **60**, 1129 (1988).

[4] G. Catalan, J. Seidel, R. Ramesh, and J. F. Scott, Rev. Mod. Phys. **84**, 119 (2012).

[5] J. Y. Jo, S. M. Yang, T. H. Kim, H. N. Lee, J.-G. Yoon, S. Park, Y. Jo, M. H. Jung, and T. W. Noh, Phys. Rev. Lett. **102**, 045701 (2009).

[6] Y. Li et al., Nat. Commun. **16**, 5451 (2025).
[7] M. Li, Z. Peng, D. Zhang, X. Wang, W. Yang, Z. Liang, and X. Jiang, Materials **19**, 1586 (2026).
[8] N. Nagaosa and Y. Tokura, Nat. Nanotechnol. **8**, 899 (2013).
[9] B. Göbel, I. Mertig, and O. A. Tretiakov, Phys. Rep. **895**, 1 (2021).
[10] H. Jani et al., Nature **590**, 74 (2021).
[11] A. Fert, V. Cros, and J. Sampaio, Nat. Nanotechnol. **8**, 152 (2013).
[12] R. E. Goldstein, D. P. Jackson, and A. T. Dorsey, Phys. Rev. Lett. **76**, 3818 (1996).
[13] A. T. Dorsey and R. E. Goldstein, Phys. Rev. B **57**, 3058 (1998).
[14] A. Cēbers, C. Gourdon, V. Jeudy, and T. Okada, Phys. Rev. B **72**, 014513 (2005).
[15] J. Ge, J. Gutierrez, B. Raes, J. Cuppens, and V. V. Moshchalkov, New J. Phys. **15**, 033013 (2013).
[16] M. Menghini and R. J. Wijngaarden, Phys. Rev. B **72**, 172503 (2005).
[17] R. Prozorov, Phys. Rev. Lett. **98**, 257001 (2007).
[18] S. Vélez, A. García-Santiago, J. M. Hernandez, and J. Tejada, Phys. Rev. B **80**, 144502 (2009).
[19] Z. Janů, Z. Švindrych, O. Truněček, P. Kúš, and A. Pleceník, Physica C **471**, 1647 (2011).
[20] S. Yonezawa and Y. Maeno, Physica C **460–462**, 551 (2007).
[21] R. A. Hein and R. L. Falge, Phys. Rev. **123**, 407 (1961).
[22] D. C. Peets, E. Cheng, T. Ying, M. Kriener, X. Shen, S. Li, and D. Feng, Phys. Rev. B **99**, 144519 (2019).
[23] H. Leng, C. Paulsen, Y. K. Huang, and A. de Visser, Phys. Rev. B **96**, 220506 (2017).
[24] F. Gömöry, Supercond. Sci. Technol. **10**, 523 (1997).
[25] X. Mi, X. Li, L. Zhang, Y. Gu, A. Wang, Y. Li, Y. Chai, and M. He, Phys. Rev. B **111**, 014417 (2025).
[26] P. Lu et al., Phys. Rev. B **113**, 224512 (2026).
[27] Y. Zhang et al., Phys. Rev. B **107**, 134417 (2023).
[28] A. Aharoni, J. Appl. Phys. **83**, 3432 (1998).

[29] T. Reimann, M. Schulz, C. Grünzweig, A. Kaestner, A. Bauer, P. Böni, and S. Mühlbauer, J. Low Temp. Phys. **182**, 107 (2016).

[30] G. Chanin and J. P. Torre, Phys. Rev. B **5**, 4357 (1972).

[31] A. N. Ulyanov, H. D. Quang, K. W. Lee, S. C. Yu, N. H. Sinh, Y. M. Kang, and S. I. Yoo, IEEE Trans. Magn. **44**, 3060 (2008).

[32] C. V. Topping and S. J. Blundell, J. Phys.: Condens. Matter **31**, 013001 (2019).

[33] H. Tamatsukuri et al., Phys. Rev. B **97**, 214407 (2018).

## Figures

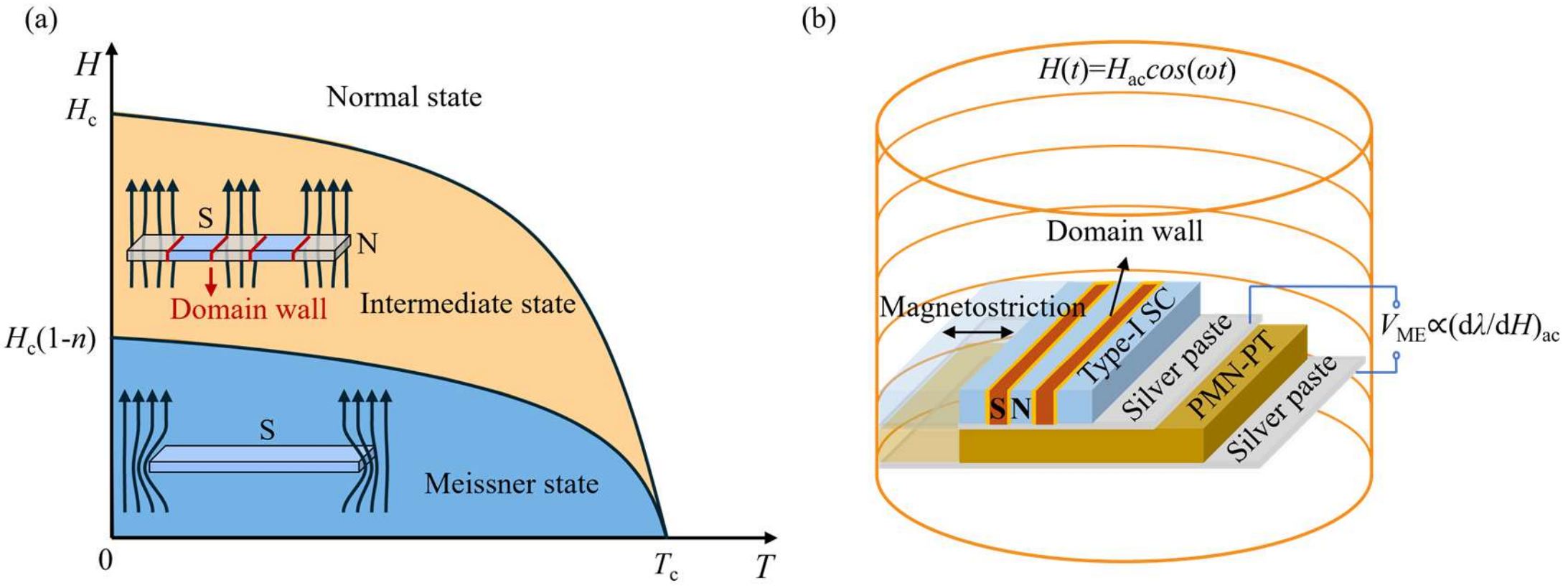

FIG. 1. Overview of type-I superconductivity and the composite ac magnetoelectric measurement setup. (a) Schematic phase diagram of a type-I superconductor. (b) Schematic of the composite ac magnetoelectric measurement setup.

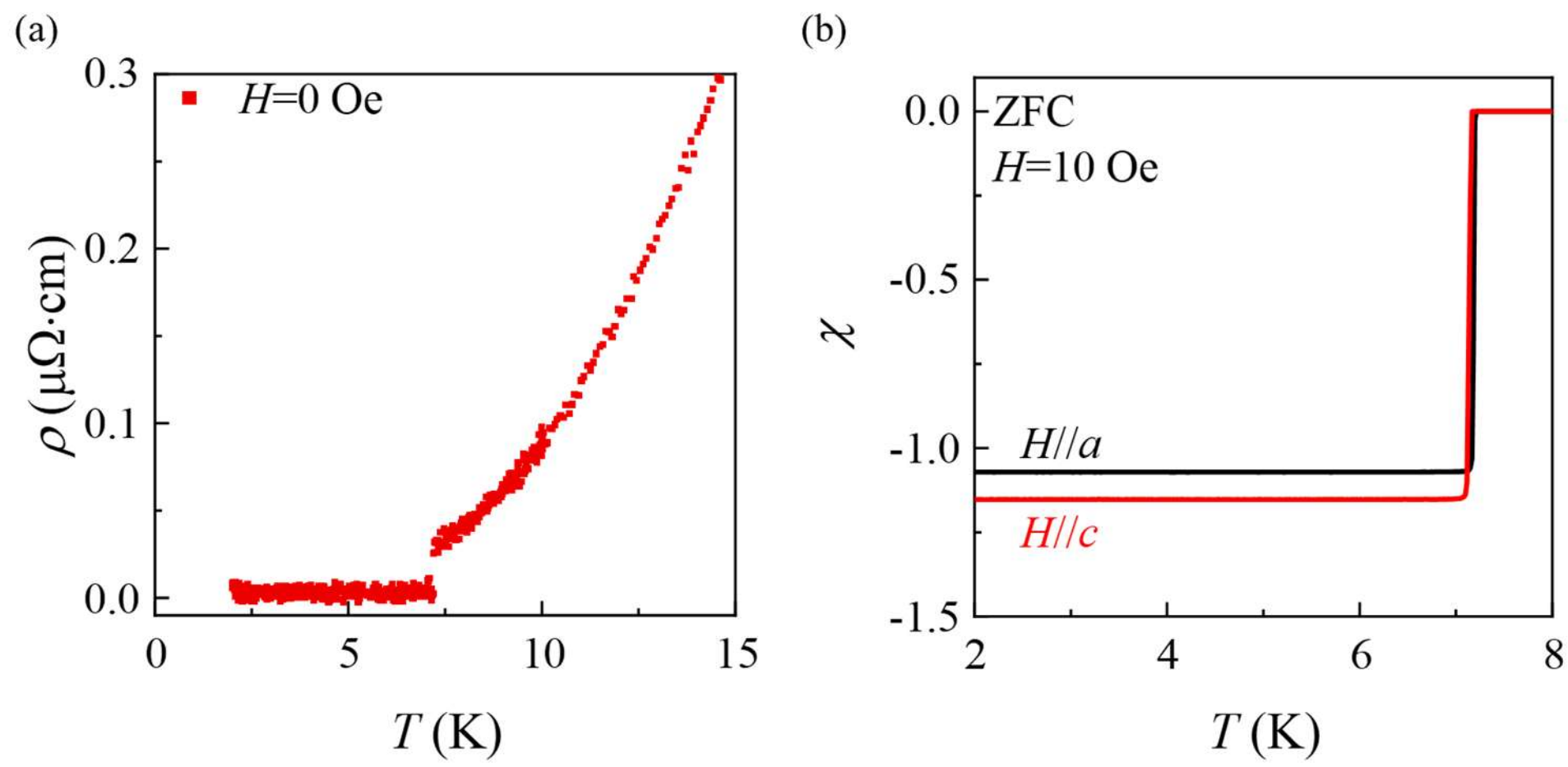

FIG. 2. (a) Temperature-dependent resistivity of Pb in zero magnetic field. (b) Superconducting volume fraction estimated from the zero-field-cooled magnetization measured at 10 Oe.

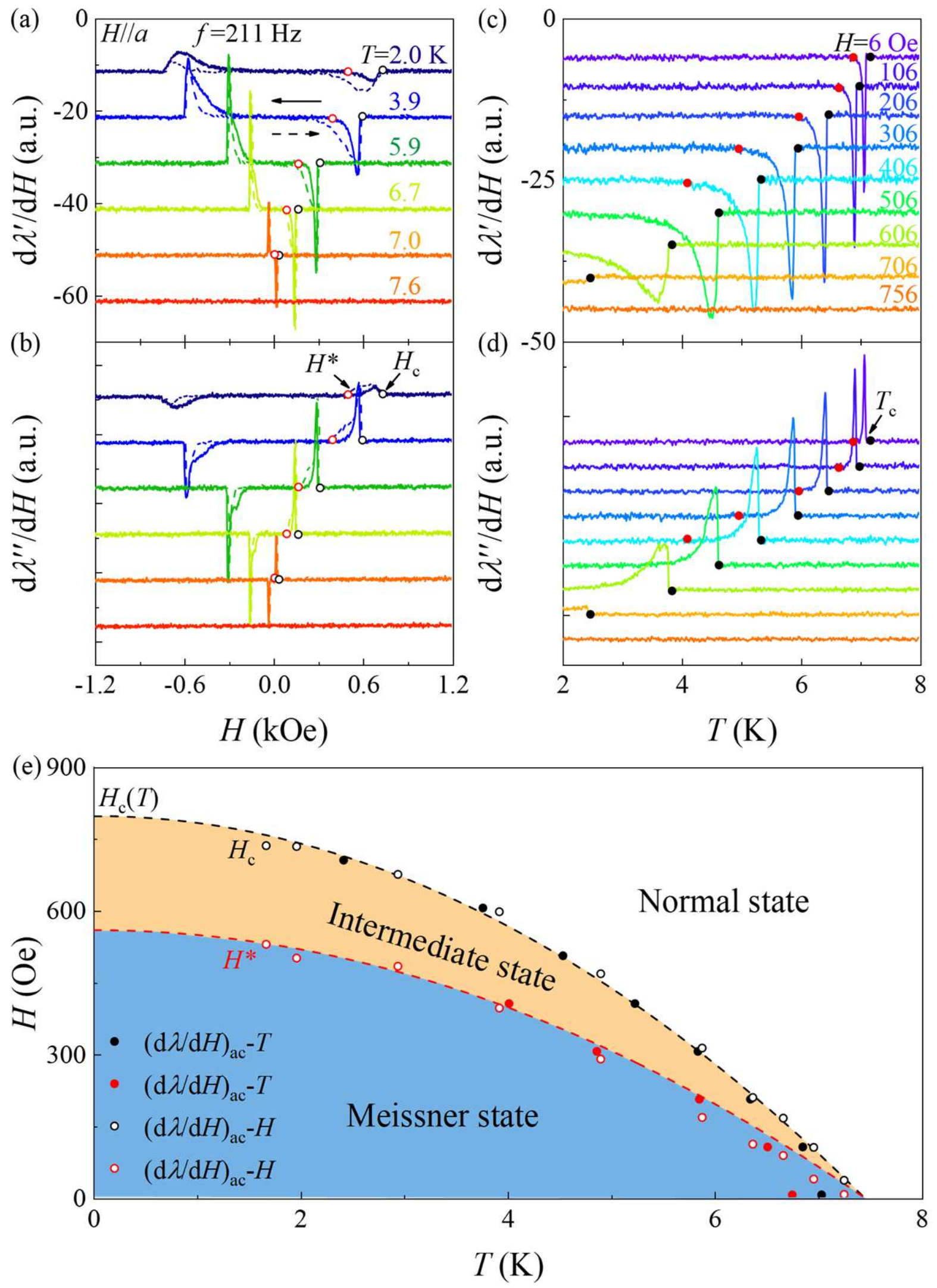

FIG. 3. Dynamic response and phase diagram of type-I Pb. (a, b) Field dependence and (c, d) temperature dependence of the real part ($d\lambda'/dH$) and imaginary part ($d\lambda''/dH$) of the ac magnetostrictive coefficient $(d\lambda/dH)_{ac}$ for $H//a$. (e) $H$-$T$ phase diagram of Pb for $H//a$. Open and solid circles mark characteristic transition points extracted from field sweeps and temperature sweeps, respectively. The black dashed line is a fit to $H_c(T) = H_c(0)[1-(T/T_c)^2]$, and the red dashed line denotes $[H_c(T)](1-n)$.

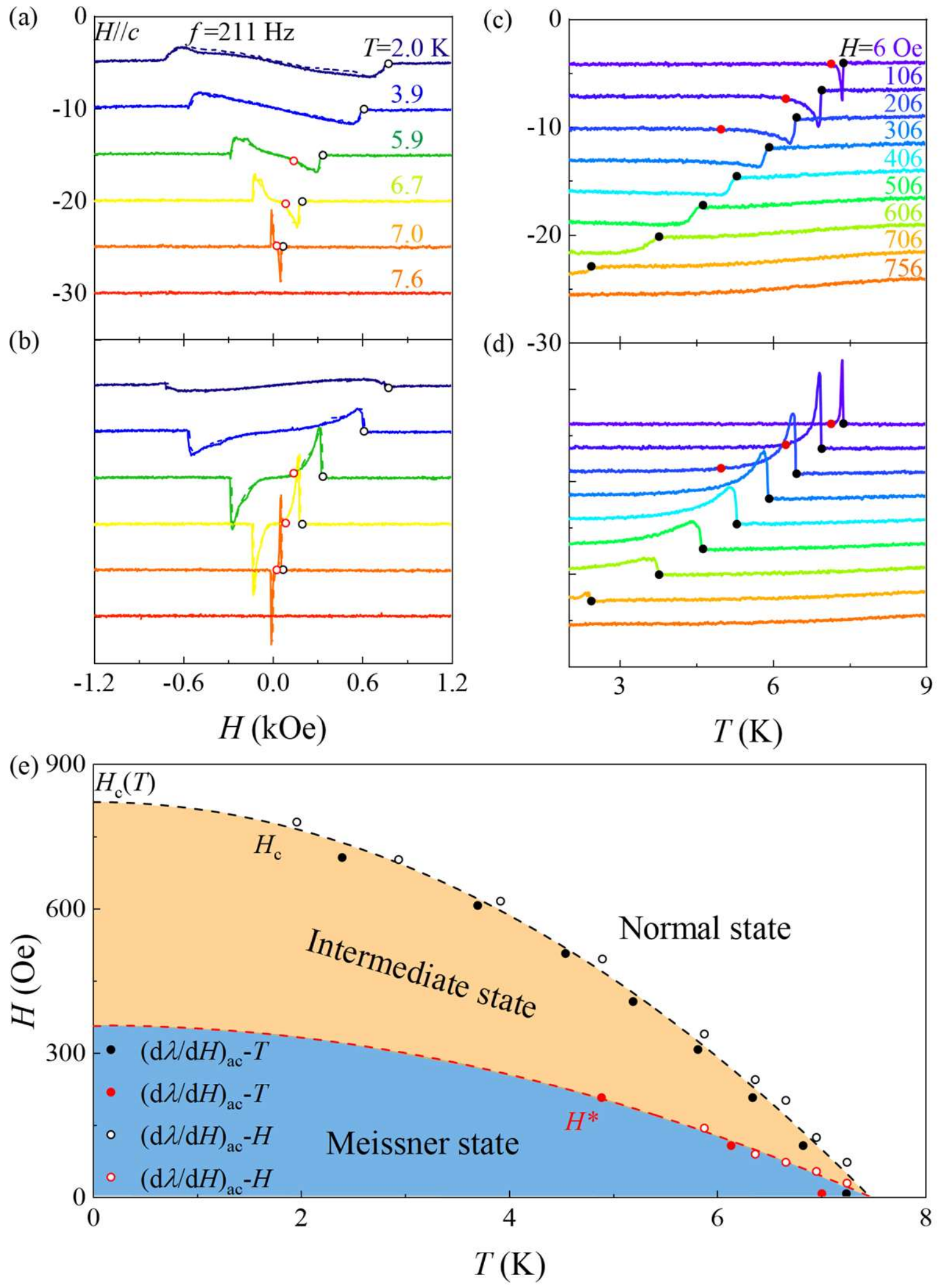

FIG. 4. (a, b) Field dependence and (c, d) temperature dependence of d$\lambda'$/d$H$ and d$\lambda''$/d$H$ for $H//c$. (e) $H$-$T$ phase diagram of Pb for $H//c$. Open and solid circles mark characteristic transition points extracted from field sweeps and temperature sweeps, respectively. The black dashed line is a fit to $H_c(T) = H_c(0)[1-(T/T_c)^2]$, and the red dashed line denotes $[H_c(T)](1-n)$.

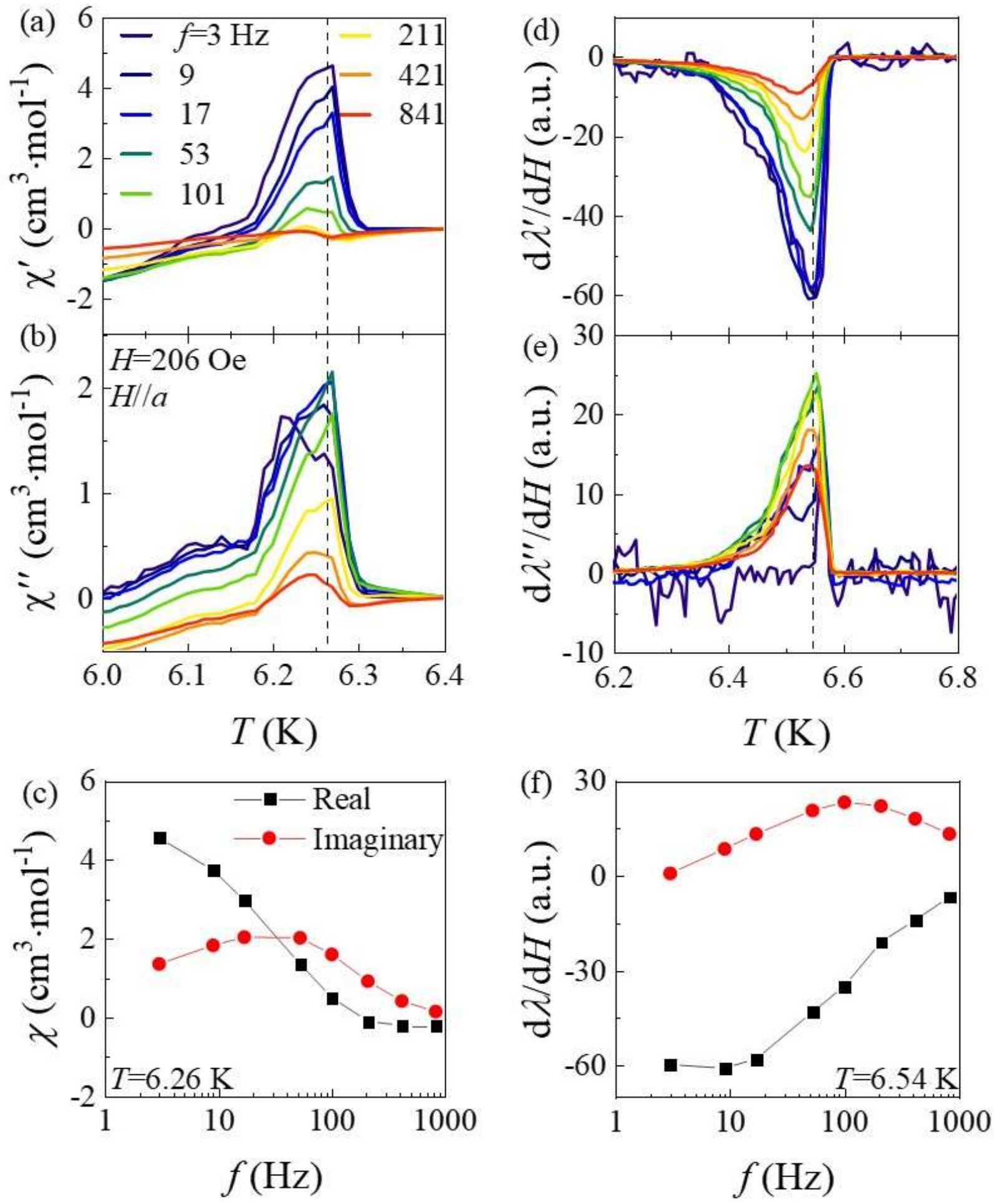

FIG. 5. Frequency-dependent contrasting behaviors of $\chi$ and $(\mathrm{d}\lambda/\mathrm{d}H)_{\mathrm{ac}}$ for $H//a$ ($H$ = 206 Oe). (a, b) Temperature dependence of the real part $\chi'$ and imaginary part $\chi''$ of the ac susceptibility $\chi$ at various frequencies. (c) $\chi'$ and $\chi''$ vs. frequency at 6.26 K. (d, e) Temperature dependence of $\mathrm{d}\lambda'/\mathrm{d}H$ and $\mathrm{d}\lambda''/\mathrm{d}H$ at various frequencies. (f) $\mathrm{d}\lambda'/\mathrm{d}H$ and $\mathrm{d}\lambda''/\mathrm{d}H$ vs. frequency at 6.54 K.

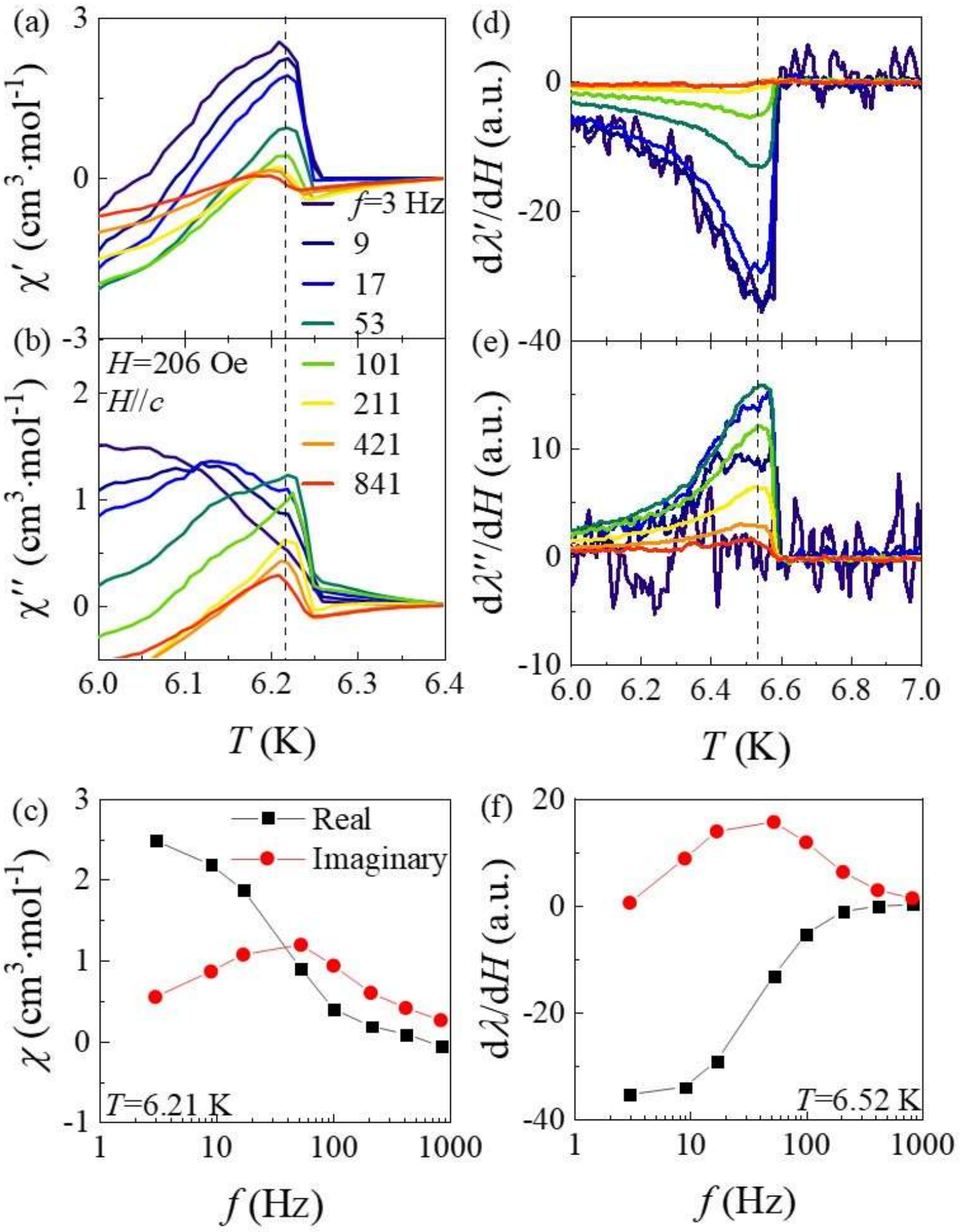

FIG. 6. Frequency-dependent contrasting behaviors of $\chi$ and $(d\lambda/dH)_{ac}$ for $H//c$ ($H$ = 206 Oe). (a, b) Temperature dependence of $\chi'$ and $\chi''$ at various frequencies. (c) $\chi'$ and $\chi''$ vs. frequency at 6.21 K. (d, e) Temperature dependence of $d\lambda'/dH$ and $d\lambda''/dH$ at various frequencies. (f) $d\lambda'/dH$ and $d\lambda''/dH$ vs. frequency at 6.52 K.

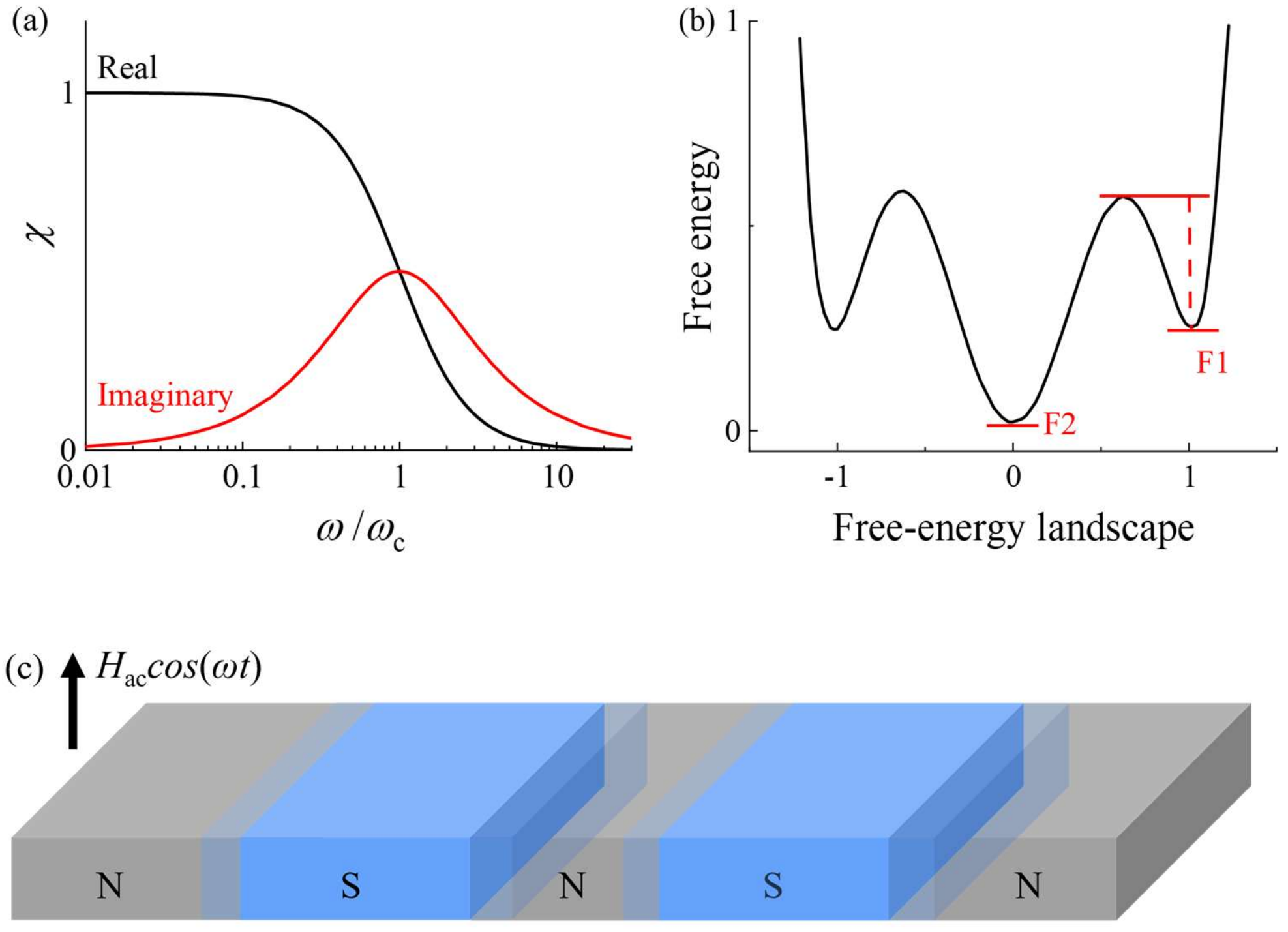

FIG. 7. (a) Theoretical curves for a Debye-type relaxation response of $\chi$. (b) Schematic free-energy landscape illustrating a metastable S/N domain configuration (F1) and a lower free energy configuration (F2) separated by a local barrier. (c) Schematic illustration of ac-field-driven S/N interface rearrangement.